\documentclass[prl,aps,showpacs]{revtex4}
\usepackage{graphicx}
\usepackage{graphics}
\begin{document}
\baselineskip=2.5ex
\title{A binuclear atom --- a special type of close bound state\\
between proton and heavy atom
}
\author{V. P. Chaly$^a$, V. L. Gurevich$^b$,  M. Yu.
Pogorelsky$^a$, Yu. V. Pogorelsky$^a$}
\affiliation{$^a$Joint-Stock Company Semiconductor Technologies
and
Equipment, P. B. 29 Saint Petersburg 194156, Russia\\
http://www.semiteq.ru, yuripogorelsky@mail.ru\\ $^b$A.  F.  Ioffe
Institute
of Russian Academy of Sciences, Saint Petersburg 194021, Russia\\
vadimlvo@hotmail.com}
\begin{abstract}
\baselineskip=3ex It is established that a bound state of a proton
with a heavy Thomas --- Fermi atom should exist. On the one hand,
the electrons of the atom screen the proton field. This decreases
the repulsion force between the proton and the nucleus. On the
other hand, the attraction force between the proton and the
electrons is directed towards the gradient of the electron
density, i.  e. towards the nucleus. For instance, for $Z=80$ both
forces become equal at approximately $0.6\,a$ where $a$ is the
Bohr radius. The corresponding minimum of the proton potential
energy is in the region of negative energies (attraction) that can
be of the order of several tens of eV. We propose to call such a
system {\em a binuclear atom}.

In contrast to the molecules where a coupling with a hydrogen atom
is due to an essential modification of one or several states of
the outer electrons the formation of a binuclear atom is a result
of collective response of the whole system of inner electrons to
the screened potential of a proton that is well inside the
electron system of the heavy atom. The variation of the wave
function of each electron can be considered as a small
perturbation. The bound state is formed as a result of joint
action of a large number (of the order of $Z$) of perturbed inner
electrons. The important problem concerning the accuracy of our
calculation is discussed.
\end{abstract}
\maketitle
\baselineskip=3ex

\section{Introduction}
The purpose of the present paper is to demonstrate that a proton
can be bound to a heavy atom with the charge of the nucleus
$Z\gg1$. The principal idea of the paper can be formulated as
follows: when a proton approaches the nucleus its potential is
screened by the atomic electrons.

As is shown by Teller~\cite{T} (see also Ref.~\cite{LS}), two
Thomas --- Fermi (TF) atoms cannot form a bound state. In the
present paper we consider what can be looked upon as an opposite
case. The principal difference of this case and the problem
considered by Teller can be formulated as follows. If one
considers two heavy atoms the Teller theorem states that such a
system has a lower energy than the same system where the atoms are
merged into a molecule. However, in the present paper we consider
a situation where one of the interacting items is a heavy atom
($Z\gg1$) that can be described by the TF theory whereas the
second item is a proton with $Z=1$. Our purpose is to prove that
in this case (not covered by the Teller's theory) a bound state
can be formed. The screened potential of a proton can be
considered as a perturbation for the electrons of the TF atom. The
bound state is formed as a result of joint action on the proton of
a large number of the perturbed inner electrons of the atom. The
vibrational energy of the proton, i. e. the energy of its
vibrations near the equilibrium position is much smaller than the
binding energy. Therefore one can consider the proton in the
S-state to be positioned on a sphere of a fixed radius.

There is a number of papers where bound states between a positron
and an atom are considered, see for instance,
Refs.~\cite{DFH,BM1,BM2} and the references therein. The ability
of positrons to bind to a number of atoms is now well established.
The attractive electron-positron interaction leads to formation of
a cluster that includes a positron {\em in the outer valence
region of the atom.}

In contrast to the problem of a bound state of a positron the
methods and results of these papers cannot be directly applied to
the situation considered here because of a great difference
between the positron and proton masses. In the binuclear atom
treated in the present paper the distance between the heavy
nucleus and the proton {\em is smaller than the Bohr radius} $a$.
It means that in this case there is a quite different physical
situation where the proton interacts practically entirely with the
{\em inner electrons of the atom.}

\section{Preliminary considerations}
To visualize the physics of this phenomenon we start with analysis
of a subsidiary problem. We will consider a Thomas
--- Fermi screening of a proton field by a homogeneous electron system
having average density $N$ neutralized by immobile positive charges. We
have
\begin{equation} \delta
N={1\over\pi^{2}}{p_Fm\over\hbar^3}e\varphi,\quad \nabla^2_{\bf
R}\varphi+q^2\varphi=0,\quad \varphi(R)={1\over R}\exp(-qR),
\label{1}
\end{equation}
$p$ is the Fermi momentum, $q$ is the reciprocal screening radius
of a proton \cite{LL10}, \S 40. Now,
\begin{equation}
q^2={4\cdot3^{1/3}\over\pi^{1/3}}{me^2\over\hbar^2}N^{1/3}=
{4\cdot3^{1/3}\over\pi^{1/3}}{1\over a}N^{1/3}.
\label{2}
\end{equation}
Here $e$ is the elementary charge, $m$ is the electron mass, $N$
is their concentration, $\delta N$ is its variation due to the
electrostatic potential $\varphi$. The energy of electrostatic
interaction of the screened proton potential situated at $R=0$
with the electron density $N$ is
\begin{equation}
E=-e^2N\int\limits_0^{\infty}d^3R{1\over R}\exp(-qR)=
-{\pi^{4/3}\over3^{1/3}}ae^2N^{2/3}. \label{3}
\end{equation}

Let us assume that $N$ is a slow (as compared with $1/q$) function
of $\bf R$: $N=N({\bf R})$. Then the electron system will act on
the proton with the force $\bf F$ given by
\begin{equation}
{\bf F}=-{\partial E\over\partial {\bf R}}=2\left({\pi\over3}\right)
^{4/3}ae^2N^{-1/3}(R) \nabla_{\bf R}N.
\label{4}
\end{equation}
Its direction is determined by $\nabla_{\bf R}N$.

\section{Calculation of the interaction potential}
Now we will treat the problem we are interested in, namely a
proton in the field of electrons and a heavy nucleus. The electron
system will be treated within the model of Thomas
--- Fermi atom (see, for instance, Ref.~\cite{LL3}, \S70;~\cite{M}).
This model was proposed long ago but quite recently an interest to
its application was revived (for instance Refs.~\cite{A,A2,A3}).
According to the model half of the electron charge of the atom is
within a sphere of the radius 1.33$aZ^{-1/3}$ where $a$ is the
Bohr radius. The electron density enhances as the distance from
the nucleus becomes smaller.  Let the distance between the nucleus
and the proton be of the order of $a$.  Its potential should be
screened by the electrons.  As a result, the repulsion between the
proton and nucleus becomes much smaller.  As the electron density
increases while $R$ decreases the electrons should attract the
proton. The force of attraction cannot be calculated using
Eq.~(\ref{4}) as $\nabla_{\bf R}N$ now is not small. The
calculations below show that for $Z=80$ the force of attraction by
the electrons and the force of repulsion by the nucleus are
counterbalanced for $R\approx0.6\,a.$ This is an indication for
the possibility of a bound state formation. We will be interested
in the bound state with the lowest energy, i.  e. the lowest
S-state.

We wish to emphasize that there is a principal difference between
the unique state treated in the present paper and an ordinary
molecule of a chemical compound containing a hydrogen atom. In
such compounds the coupling of a hydrogen atom is due to an
essential modification of one or a few outer electron states
whereas here we consider a special effect that is a collective
change of the whole distribution of the atom's inner electrons by
the proton potential. In ordinary molecules a coupling with a
hydrogen atom is due to an essential modification of one or
several states of the outer electrons. In the present paper we
consider a special effect that is a collective response of the
whole system of inner electrons to the screened potential of a
proton. In regard to each electron the action of the potential can
be considered as a small perturbation. The bound state is formed
because these perturbations are summed up. Thus the perturbation
theory is an adequate approach to this problem. Usually the
perturbation theory cannot describe a bound state formation.
However, the present case is unique as the bound state is formed
as a result of united action of a large number of perturbed
electrons.

Consider the screening of a proton by the electrons of a heavy
atom. We assume the distance between the proton and nucleus to be
$\lesssim a$. In other words, the proton is well inside the atom.
This is why we propose to call such a system {\em a binuclear
atom}. We will make estimates of the energy of such system within
the TF atom model. In the first order of the perturbation theory
the variation of electron density $\delta N$ is
\begin{equation}
\delta N({\bf r})=4\sum_i^{i_{\rm
max}}\sum_k^{\infty} \langle k\vert\varphi\vert i\rangle
{1\over\varepsilon_i-\varepsilon_k}Y_{lm}(\theta,\chi)
Y_{l'm'}(\theta,\chi)\psi_{nl}(r)\psi_{n'l'}(r) .
\label{5}
\end{equation}
Here $i=nlm$ are the occupied states, $k=n'l'm'$, $n\neq n'$
and/or $l\neq l'$; $\theta, \chi$ are the spherical angles,
$\psi_{nl}$ is the radial quasiclassical (WKB) electron wave
function in the potential $\Phi(r)$ of the TF atom. Here and
henceforth we use the atomic units $e=\hbar=m=1$.

Let us transform Eq.~(\ref{5}) into such a form that it gives a
local relation between $\delta N$ and $\varphi$. $\delta N$ being
given, $\varphi$ is determined by the Poisson equation.
Inserting the relation between $\delta N$ and
$\varphi$ we get a selfconsistent equation for $\varphi$.

The variation of the electron energy in the first order of the
perturbation theory is
\begin{equation}
\delta\varepsilon_i=\langle i\vert\varphi\vert i\rangle.
\label{6}
\end{equation}
The perturbation for every electron is the proton potential plus
the potential produced by $\delta N$, i. e. the screened proton
potential. In (\ref{5}) and (\ref{6}), there is a contribution of
a selfacting electron. It gives a contribution to the screened
potential $\varphi$. This contribution is however negligibly small
as the electron number $Z$ is large. The total energy variation of
the system $\Delta E_1$ is the sum of expressions (\ref{6}) over
all the particles including the nucleus

\begin{equation}
\Delta E_1=Z\varphi(0)-\int N({\bf r})\varphi({\bf
r})d^3r.
\label{7}
\end{equation}

Let $z$-axis join the nucleus and the proton. Then $\varphi({\bf
r})$ does not depend on the azimuthal angle and can be expanded as
\begin{equation}
\varphi({\bf r})=\sum_s\varphi_s(r)P_s(\cos\theta)   \label{8}
\end{equation}
where $P_s$ is a Legendre polynomial.
As $N(r)$ has a spherical symmetry, only $\varphi_0(r)$ contributes
to Eq.~(\ref{7}). Now we will expand $\delta N$ like Eq.~(\ref{8})
and integrate Eq.~(\ref{5}) over $\theta$
using the well-known identity
\begin{equation}
\sum_{m=-l}^l|Y_{lm}|^2=(4\pi)^{-1}(2l+1).
\label{9}
\end{equation}
As a result, Eq.~(\ref{5}) takes the form
\begin{equation}
\delta N_0(r)=\pi^{-1}\sum_l(2l+1)\sum_{n'}^{{\infty}}\sum_n^{n_{\rm
max}} \langle n'\vert\varphi_0\vert n\rangle{{1\over\varepsilon_{nl}
-\varepsilon_{n'l}}}\psi_{nl}(r)\psi_{n'l}(r).
\label{10}
\end{equation}

The radial quasiclassical function of the state ($n,l$) with
the energy near the Fermi level is
\begin{equation}
\psi_{nl}(r)=a_{nl}r^{-1}p_l^{-1/2}\cos\left[\int_{r_1}^r
(p_l+\varepsilon_{nl}p_l^{-1})dr'+C_l\right], \quad
p_l^2=2\Phi_l(r)\equiv2\Phi(r)-{1\over r^{2}}\left(l+
{1\over2}\right)^2
\label{11}
\end{equation}
($C_l$ are constants of the order of 1).
In Eq.~(\ref{11}) the approximation $\left(p_l^2+2
\varepsilon_{nl}\right)^{1/2}\approx p_l+\varepsilon_{nl}p_l^{-1}$,
is used. It is valid for $\vert\varepsilon_{nl}\vert\ll\bar{\Phi}_l$
where $\bar{\Phi}_l$ are the characteristic values of ${\Phi}_l(r)$,
$a_{nl}^2=(2/\pi)d\varepsilon_{nl}/dn=4T_l^{-1}$, $T_l$ is
the period of the classical motion at the Fermi level,
$r_1$ (as well as $r_2$) are the classical turning points.

Introducing the variable
$$
t_l(r)=\int\limits_{r_1}^r[p_l(r')]^{-1}dr', \quad 2t_l(r_2)=T_l,
$$
we expand $\varphi_0(r)$ into Fourier series
\begin{equation}
\varphi_0(r)=\sum_{\nu=0}^{\infty}\varphi_{0\nu}\cos(2\pi \nu T_l^{-1}t_l).
\label{12}
\end{equation}
Discarding the quickly oscillating part of the product of functions
given by Eq.~(\ref{11}) one gets
\begin{equation}
\delta
N_0(r)=\sum_l{2l+1\over 2\pi^2r^2p_l}
\sum_{\nu=0}^{\infty}\varphi_{0\nu}\!\int\limits_{-\Phi_l(r)}^{\infty}
\!d\varepsilon'\!
\int\limits_{-\Phi_l(r)}^{0}\!d\varepsilon{\cos[(\varepsilon'-\varepsilon)t_l]
\over\varepsilon'-\varepsilon}
\left[\delta\left(\varepsilon-\varepsilon'+{2\pi\nu\over T_l}\right)+
\delta\left(\varepsilon-\varepsilon'-{2\pi\nu\over T_l}\right)\right].
\label{13}
\end{equation}

The integral over $\varepsilon'$ can be presented as a sum of
integrals over the intervals $[-\Phi_l,0]$ and $[0, \infty]$. The
contribution of the first integral vanishes as the change of the
variables $\varepsilon{\leftrightarrow} \varepsilon'$ changes the
sign of this expression. The conditions $\varepsilon\leq0$ and
$\varepsilon'\geq0$ are fulfilled for the first $\delta$-function
and $\varepsilon\in[-2\pi\nu T_1^{-1},0].$ In the expansion
Eq.~(\ref{12}) only the harmonics where $2\pi\nu
T_l^{-1}\ll\bar{\Phi}_l$ are physically important. Integrating
over $\varepsilon'$ with regard of $\delta$-function and over
$\varepsilon$ in the interval $2\pi\nu T_l^{-1},0$, and taking
into account Eq.~(\ref{12}) one gets
\begin{equation}
\delta N_0(r)={1\over 2\pi^2r^2}\sum_l{2l+1\over
p_l}\varphi_{0}(r).
\label{14}
\end{equation}
$r$ is in the classical region for the quantum numbers $l$
satisfying the condition $\Phi_l(r)\geq0$. Integration over
$l$ from 0 up to the maximal value, corresponding to
$\Phi_l(r)\geq0$ gives
\begin{equation}
\delta
N_0(r)=\pi^{-2}p(r)\varphi_{0}(r), \quad p=(2\Phi)^{1/2}.
\label{15}
\end{equation}

Eq.~(\ref{15}) coincides with the first equation~(\ref{1}). Thus
the relation between $\delta N$ and $\varphi$ is preserved also
for nonhomogeneous electron density. Eq.~(\ref{15}) gives the
Poisson equation
\begin{equation} {1\over r}{d^2\over
dr^2}(r\varphi_{0})=q^2\varphi_{0},\quad q^2=4\pi^{-1}\sqrt{2\Phi}
\label{16}
\end{equation}
with the boundary conditions
\begin{equation}
\quad \left.r\varphi(r)\right\vert_{r\to0}=0,\;
\left.\varphi(r)\right\vert_{r\to \infty}=0, \quad{d\over
dR}(r\varphi_0)\left\vert_{\phantom{A}\atop{}
\displaystyle{r=R-0}}- {d\over
dR}(r\varphi_0)\right\vert_{\displaystyle{r=R+0}}={1\over R}.
\label{17} \end{equation} The jump of the derivative is due to the
presence of the proton at the distance $R$ from the nucleus. For
small values of $r$ we solve these equations numerically. For
$r>R$ one can use the eikonal approximation that is valid at
$d\Phi/dr\ll 2p^2q$. Using the TF variable $x=\gamma
R,\,\gamma=Z^{1/3}/0.885$ one can see that these solutions (but
not the last boundary condition at $x=\gamma R$) are independent
of Z. For $x<\gamma R \quad x\varphi_0(x)=CP(x)$, $P(x)$ is given
in Table 1.

\bigskip
\centerline{Table 1}

\begin{center}
\begin{tabular}[b]{||c|c||c|c||c|c||}
\hline\hline
$x$ &$P(x)$&$x$ &$P(x)$&$x$ &$P(x)$\\
\hline \hline\phantom{mmm}0.2\phantom{mmm} &\phantom{mmm} 0.207
\phantom{mmm}&\phantom{mmm}1.4\phantom{mmm}&\phantom{mmm}2.25\phantom{mmm}
&\phantom{mmm}2.6\phantom{mmm}&\phantom{mmm}6.71\phantom{mmm}\\
\hline
0.4&0.437&1.6&2.79&2.8&7.78\\ \hline 0.6&0.703&1.8&3.40&3.0&8.97\\
\hline 0.8&1.01&2.0&4.09&3.2&10.3\\ \hline
1.0&1.37&2.2&4.87&3.4&11.7\\ \hline
1.2&1.78&2.4&5.74&3.6&13.3\\
\hline \hline
\end{tabular}
\end{center}
For $x\to0, P=x$. For $x>\gamma R$
\begin{equation} x\varphi_{0}(x)=bq_x^{-1/2}\exp\left(-\int
\limits_{\gamma R}^x q_x dx\right),\quad q_x^2=1.5\cdot x^{-1/2}X^{1/2},
\quad\Phi=ZX/r.
\label{18}
\end{equation}
The constants $c$ and $b$ are determined by continuity of function
$x\varphi_0$ and the jump of derivative of function (\ref{17})
at $x=\gamma R$.

Making use of the well known relation for the Thomas --- Fermi
atom between $N(r)$ and $\Phi(r)$ and the values of $\Phi$ given
in the tables one can calculate $\Delta E_1$ Eq.(\ref{7}) for
various $R$. Note that an additional correction is needed (that,
however, will be small as compared with $\Delta E_1$). We have
considered the screening in a neutral atom. The total electron
charge surplus near the proton due to the screening is $-1$.
Indeed, the total potential of the proton and $\delta N$
Eq.~(\ref{18}) tends to zero at $r\to\infty$ faster than $r^{-1}$.
The corresponding correction $\Delta E_2$ should not exceed the
energy of one-electron ionization of the atom. In the TF atom
model this energy appears to be lower as compared to the
experimental values. Such a discrepancy is due to the fact that in
this model such properties are essentially related to the
distances $r\gg1$ where the TF approximation fails. The average
(over a number of different atoms) experimental value of the
ionization energy is about 0.3 atomic units. The variation of the
energy $\Delta E=\Delta E_1+\Delta E_2$ is presented in Fig. 1.
Here we have assumed $\Delta E_2=0.3$~a. u.

Let us analyze the approximations used to obtain $\Delta E$.
First, this is the one-electron approximation. The characteristic value
of the kinetic energy of the electrons should be bigger than the
potential energy of their Coulomb interaction, i. e.
$p^2/2\gg N^{1/3}$. For the atoms where $Z\gg1$ this condition is
satisfied. Second, this is the quasiclassical approximation. It is
satisfied in the interval $Z^{-1}<r<1$. The solution (\ref{18})
decays exponentially for $r>R$. This is why the integral (\ref{7})
is determined by the interval $r\lesssim R$.

We have calculated the energy $\Delta E$ for $Z$=80 and $R<1$. For
$R>1$ our theory ceases to be applicable. [When calculating the
integral~(\ref{7}) the interval of integration has been taken from
0 to 1 where the integrand is well defined]. $N(r)$ decreases with
$r$; $\varphi_0(r)$ increases for $r<R$ and decreases for $r>R$.
The integrand varies slowly for $r<R$ and rapidly decays for
$r>R$. Thus for $R<0.8$ the discarded additional energy is small.

\begin{figure}
  \includegraphics[width=4.4in]{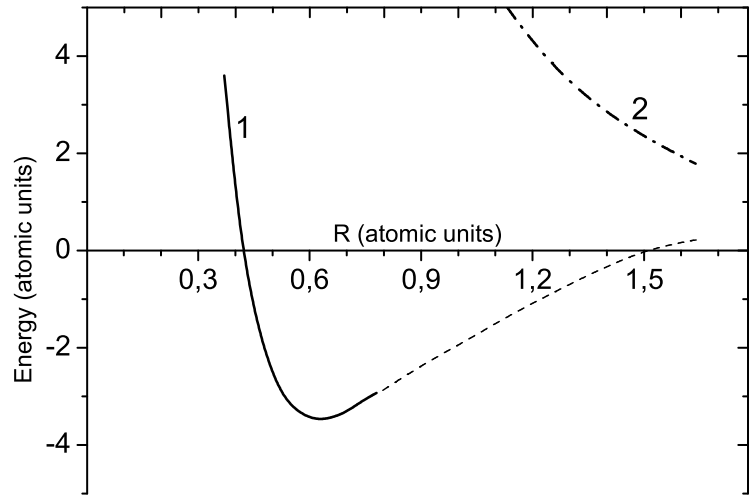}\\
  \caption{Fig. 1: 1 --- the energy of the system proton --- heavy atom
($Z$=80) as a function of the distance $R$ between the proton and
the nucleus. 2 --- the potential $\Phi(R)$. For $R>1$ one can
expect that curve 1 will tend to $\Phi(R)$. A possible transition
curve is indicated by the broken line.}\label{fig2}
\end{figure}

Fig. 1 shows that the energy of the system proton --- heavy atom
($Z\approx80$) with regard of screening appears to be negative at
$R>0.4$ a. u. and has a minimum at $R\approx0.6$ a. u.($\approx$
0.3 \AA).  One can expect that at the periphery of the atom
$\Delta E$ can become positive (potential barrier).

The distance between the bound proton and the nucleus ($\approx$
0.3 \AA\, for $Z\approx\,$80) is much smaller than the
characteristic radius of the valence electron wave functions. This
means that the chemical behavior of such a system after an extra
electron has joined it should be similar to that of an atom of the
number $Z$+1.

\section{Conclusion}
The experimental search of the predicted phenomenon should be
probably performed with plasmas or using a source of low-energy
protons. The energy release in the course of proton trapping might
be accompanied by Auger-processes and X-ray emission. The
probability of a proton capture might be bigger for a heavy atom
in a condensed matter (i. e. for a target) than for a vapor. The
described method is applicable for estimates of the binding energy
of two protons with a heavy atom. If the average distance between
the protons is of the order of 0.3 \AA\, the Coulomb energy of
their repulsion is smaller that the coupling energy obtained in
the present paper. Our estimates indicate that a possibility
exists for trapping of $\alpha$-particles and $\mu^+$-mesons.

The TF model describes the average atomic characteristics. It
cannot describe their individual properties, such as their
periodicity. This point concerns also our estimates.

A few words about the accuracy of our calculation. In general the
accuracy one can expect of the TF atom model is about 10 --- 15
per cent. Calculating the potential curve one deals with a
difference of two large quantities. This means that the difference
of these quantities should be bigger than 15 per cent of each of
them. Near the minimum of the potential curve these two quantities
differ by 30 per cent. This means that the accuracy of our
calculation is sufficient. However, it is desirable to apply in
future more advanced methods to obtain a more detailed value for
the potential minimum position as well as the form of potential
curve near the minimum. This should facilitate the experimental
search of the binuclear atoms.

\vspace{1cm}
\centerline{Figure caption}

\bigskip
Fig. 1. 1 --- the energy of the system proton --- heavy TF atom
($Z$=80) as a function of the distance $R$ between the proton and
the nucleus. 2 --- the potential $\Phi(R)$. For $R>1$ one can
expect that curve 1 will tend to $\Phi(R)$. A possible transition
curve is indicated by the broken line.
\end{document}